# Experimental Demonstration of a Reconfigurable Coupled Oscillator Platform to Solve the Max-Cut Problem

Mohammad Khairul Bashar[1*], Antik Mallick[1*], Daniel S Truesdell[1], Benton H. Calhoun[1], Siddharth Joshi[2], and Nikhil Shukla[1]

[1]Department of Electrical & Computer Engineering, University of Virginia, Charlottesville, VA-22904, USA,
[2]Department of Computer Science and Engineering, University of Notre Dame, USA-46556,
email: ns6pf@virginia.edu, *equal contribution

This work was supported in part by Semiconductor Research Corporation (SRC), under task 2841.001 and by National Science Foundation (NSF) ECCS-1807551.

*Abstract*— **In this work, we experimentally demonstrate an integrated circuit (IC) of 30 relaxation oscillators with reconfigurable capacitive coupling to solve the NP-Hard Maximum Cut (Max-Cut) problem. We show that under the influence of an external second-harmonic injection signal, the oscillator phases exhibit a bi-partition which can be used to calculate a high quality approximate Max-Cut solution. Leveraging the all-to-all reconfigurable coupling architecture, we experimentally evaluate the computational properties of the oscillators using randomly generated graph instances of varying size and edge density (η). Further, comparing the Max-Cut solutions with the optimal values, we show that the oscillators (after simple post-processing) produce a Max-Cut that is within 99% of the optimal value in 28 of the 36 measured graphs; importantly, the oscillators are particularly effective in dense graphs with the Max-Cut being optimal in seven out of nine measured graphs with η=0.8. Our work marks a step towards creating an efficient, room-temperature-compatible non-Boolean hardware-based solver for hard combinatorial optimization problems.**

## I. Introduction

Digital computing has been the backbone of modern information processing technology. Despite its tremendous strides, there is a class of computational problems, commonly referred to as NP-Hard problems, that are still considered fundamentally intractable to compute using digital computers. A case in point, and the focus of the present work, is computing the maximum cut (Max-Cut) of a (unweighted) graph G(V,E) (V: vertices; E: edges) which is a cut that divides G into two sets such that the number of common edges between them is as large as possible; the number of common edges is the value of the Max-Cut. The Max-Cut problem is an archetypal NP-Hard problem [1] that finds extensive use in areas ranging from statistical physics [2]-[4], medicine discovery to VLSI design [5]. However, solving the problem using conventional digital computers entails an exponential increase in computational resources as the size of the problems increase. Subsequently, this has motivated the search for alternate computing platforms [6]-[11], such as Ising machines (based on the Ising model) evaluated here, that can potentially provide a more efficient pathway to solving such problems. It is worth emphasizing that the successful realization of such a non-Boolean platform (e.g. the coupled oscillators explored here) is likely to benefit the broader class of such problems since many such problems can be formulated in terms of the underlying Ising model [9] through polynomial time transformations.

In this work, we experimentally develop an integrated circuit (IC) of CMOS-compatible coupled relaxation oscillators and demonstrate its functionality as an Ising machine to compute the Max-Cut of a graph. The Max-Cut problem can be mapped directly to an Ising Hamiltonian: $H = -\sum_{i,j}^{N} J_{ij}\sigma_i\sigma_j$, where each spin σ corresponds to a node of the graph and can take binary values $\sigma \in \{\pm 1\}$, N is the total number of nodes in the problem, and $J_{ij}$ is interaction coefficient between nodes *i* and *j*. Computing the Max-Cut solution then corresponds to minimizing $H$ [12]. Consequently, there has been an active research effort to realize a physical 'Ising machine' that inherently evolves to minimize its energy, and thus, naturally computes the Max-Cut solution. Examples of such demonstrations include the D-Wave quantum annealer [13]-[15], optical parametric oscillator-based Coherent Ising machines (CIM) [16]-[18], and SRAM-based Ising machines that use CMOS annealing [19], as well as the new CMOS annealing processors that use processing- in-/near- memory [20]-[22]. Coupled oscillators have also been explored as an alternate non-Boolean approach to solving computationally hard problems [23]-[35], and more importantly, have recently been shown to behave as Ising machines [36]-[39] relevant to solving the Max-Cut problem. Wang *et al.* [36], [37] and Chou *et al.* [38] recently demonstrated Ising machines using resistively coupled sinusoidal oscillators operating under the influence of a second harmonic injection signal, and Dutta *et al.* [39] showed a similar functionality in four capacitively coupled injection-locked $VO_2$ oscillators. Furthermore, Ahmed *et al.* [40] recently demonstrated a scaled integrated circuit (IC) of 560 hexagonally connected CMOS-based ring oscillators to solve the Max-Cut in large planar graphs. These works attest to the increasing interest in exploring coupled oscillators to solve computationally challenging problems.

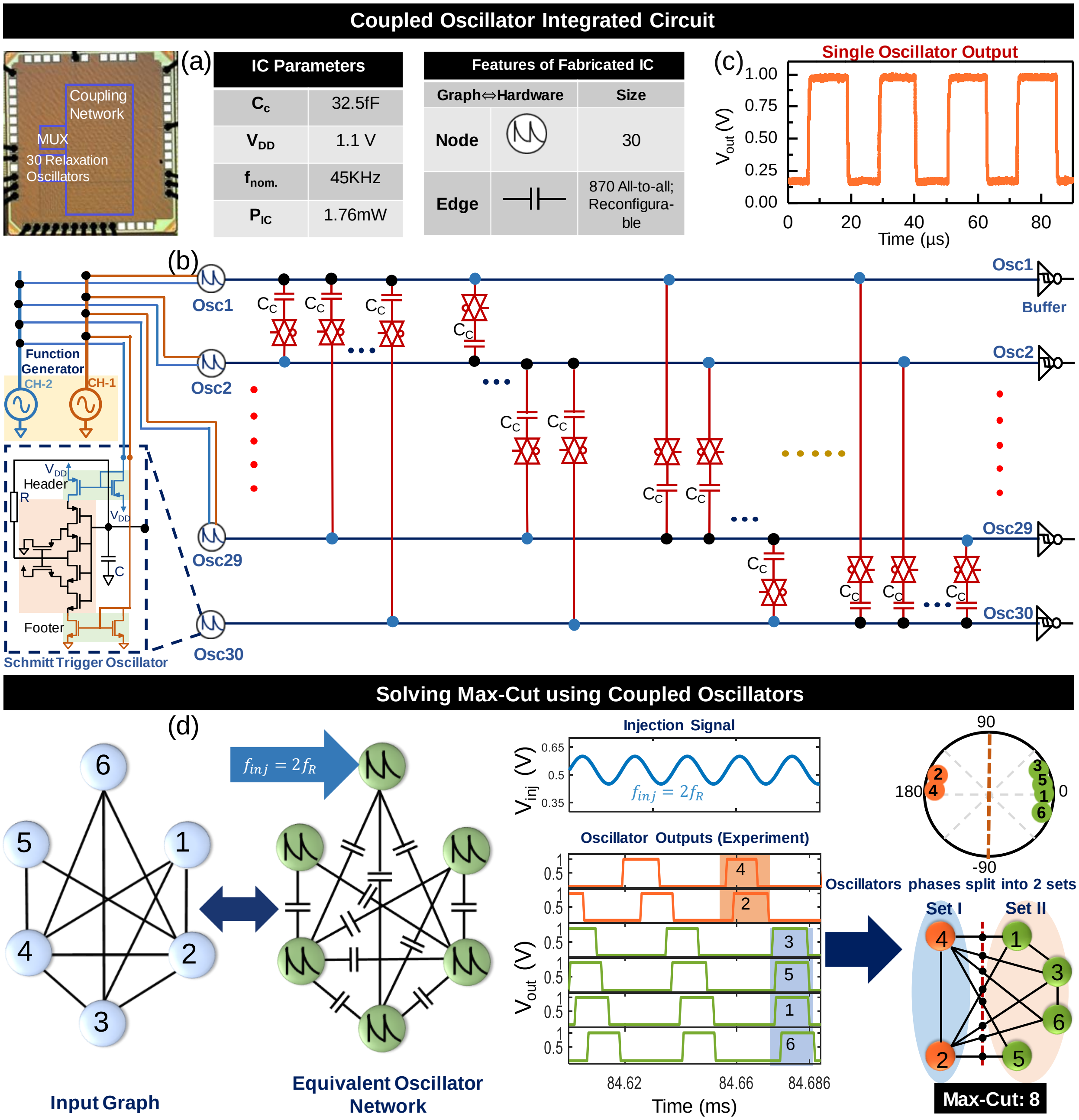

Fig. 1. (a) Die photograph and the operating parameters of the fabricated IC. (b) Circuit schematic of the coupled oscillator circuit. The oscillator is implemented using an invertering Schmitt-trigger design with feedback. The programmable coupling element consists of a capacitor in series with a T-gate. The coupling architecture enables each oscillator to be coupled to any and all other oscillators in the circuit; (c) Output of the free running relaxation oscillator. (d) Experimentally measured dynamics of the coupled oscillators (operating under the influence of a second-harmonic injection locking signal) for a representative 6 node graph. It can be observed from the polar plot that the oscillator phases exhibit a bi-partition which can be used to compute a high quality Max-Cut of the graph.

Here, we demonstrate a coupled relaxation oscillator IC to solve the Max-Cut problem (Fig. 1) in non-planar graphs. Our platform incorporates: (a) 30 programmable CMOS Schmitt-trigger-based relaxation oscillators that operate under the influence of a second-harmonic injection signal ($f_{inj} \cong 2*f_R$ where $f_R$ is the resonant frequency); (b) Reconfigurable and symmetric capacitive coupling among the oscillators i.e. any oscillator can be coupled to any and all other oscillators which allows us to process a graph (up to 30 nodes) with arbitrary connectivity.

Subsequently, we show that by creating a network that is topologically equivalent to the input graph i.e. each oscillator is mapped to a node of the graph and every coupling capacitor corresponds to an edge, the resulting phase dynamics of the oscillators can be empirically used to compute an approximate Max-Cut solution. The oscillator phases exhibit a bi-partition i.e. $0^o$ or $180^o$ which corresponds to the two sub-sets created by the Max-Cut. We note that the external sub-harmonic signal helps induce the bi-partition relevant to the Max-Cut; without this signal, the oscillators exhibit a continious phase ordering (as shown in our prior work [26] and also observed in this IC but not shown here). The advantage of the developed hardware is that besides being compatible with state-of-the-art CMOS foundry processes, it is compact (unlike other Ising implementations such as CIM) [41] and suited for room temperature operation (unlike Quantum Annealing) [42]. Furthermore, the reconfigurability incorporated in the design gives us a unique opportunity to characterize and evaluate the dynamics and the computational properties of the system over a range of graph sizes and connectivity.

## II. COUPLED OSCILLATOR IC

The coupled oscillator IC is fabricated using bulk CMOS 65nm node technology (Fig. 1(a)). Each oscillator is implemented using a Schmitt trigger inverter with a negative RC feedback (Fig. 1(b)); the feedback resistor is implemented using a switched capacitor. Furthermore, current mirrors are implemented at the header and the footer of each oscillator to control the oscillation frequency, and importantly, also enable injection of the second-harmonic signal. The external injection signal is a sinusoidal signal with a peak-to-peak amplitude of 150mV, DC offset (footer: 0.5 V, header: 0.3 V), and a frequency ($f_{inj}$) approximately twice the resonant frequency of the coupled circuit, $i.e.$ $f_{inj} \cong 2f_R$. This signal is generated using an external function generator (two separate channels were used to achieve the different DC offsets required for the header and footer) and injected to the header and the footer circuit. To sense the resonant frequency, the coupled oscillator circuit (corresponding to the graph) is first measured without the injection signal. Each oscillator output is buffered & digitized using a hysteretic Schmitt-trigger buffer to facilitate read-out while preserving the phase information. Fig. 1(c) shows the output from a single oscillator. The coupling architecture shown in Fig. 1(b) is implemented as a 30 line bus wherein an oscillator can be coupled to any and all other oscillators through the bus using a capacitor in series with a T-gate which is used to program (ON/OFF) the coupling between any two oscillators; the coupling capacitor is implemented as a metal-insulator-metal capacitor with an area of 14.78um$^2$ and having a value of 32.5 fF. This value of coupling capacitance was chosen since it was in the range of values wherein the system demonstrated the desired phase dynamics. We observed using simulations that for very small values of coupling capacitance (<5fF), the oscillators may fail to lock for certain coupling configurations. In contrast, for large capacitances (>120 fF), the system exhibited inphase locking for certain graphs. A total of 870 coupling elements enable programmable and symmetric coupling between any and all other oscillators in the network. Serial-In-Parallel-Out registers are used to program the oscillators and the coupling elements; a 32:1 MUX is used for serial read-out. The output of one of the oscillators is also tapped directly (besides passing through the MUX) and serves as the reference to which the phases of the other oscillators are compared. The power dissipated in the chip is measured to be 1.76 mW.

## III. RESULTS

### A. *Computing Max-Cut using Coupled Oscillators*

To compute the Max-Cut of a graph using the coupled oscillators, we start with the adjacency matrix A of the graph where $A_{ij}$ indicates the presence ($A_{ij}$=1) or absence ($A_{ij}$=0) of an edge between node *i* and *j* of the graph. Since we consider undirected graphs here, $A_{ij}=A_{ji}$. Each node of the graph is mapped to an oscillator and every edge (represented by $A_{ij}=A_{ji}$) to a coupling capacitor; node ≡ oscillator; edge ≡ coupling capacitor; oscillator phase ≡ set (created by the cut) to which the node belongs. In the context of the proposed hardware, the number of rows (or columns) in A represents the number of oscillators required to process the graph, and $A_{ij}$ is used to configure the corresponding coupling among the oscillators. The capacitors couple the oscillators negatively [39] i.e. oscillators exhibit phase repulstion when capacitively coupled, and have a negative relationship to the edge weight. The matrix A is passed on to the SIPO registers to initialize a topologically similar oscillator network. Fig. 1(d) shows the experimentally measured oscillator outputs along with the corresponding phase plot for a representative 6 node graph. A bi-partition in the oscillator phases observed in the polar plot corresponds to the two subsets (Set I and II) created by the (Max-)Cut; the Max-Cut value can subsequently be computed by counting the number of common edges (=8 in this example) between the sets.

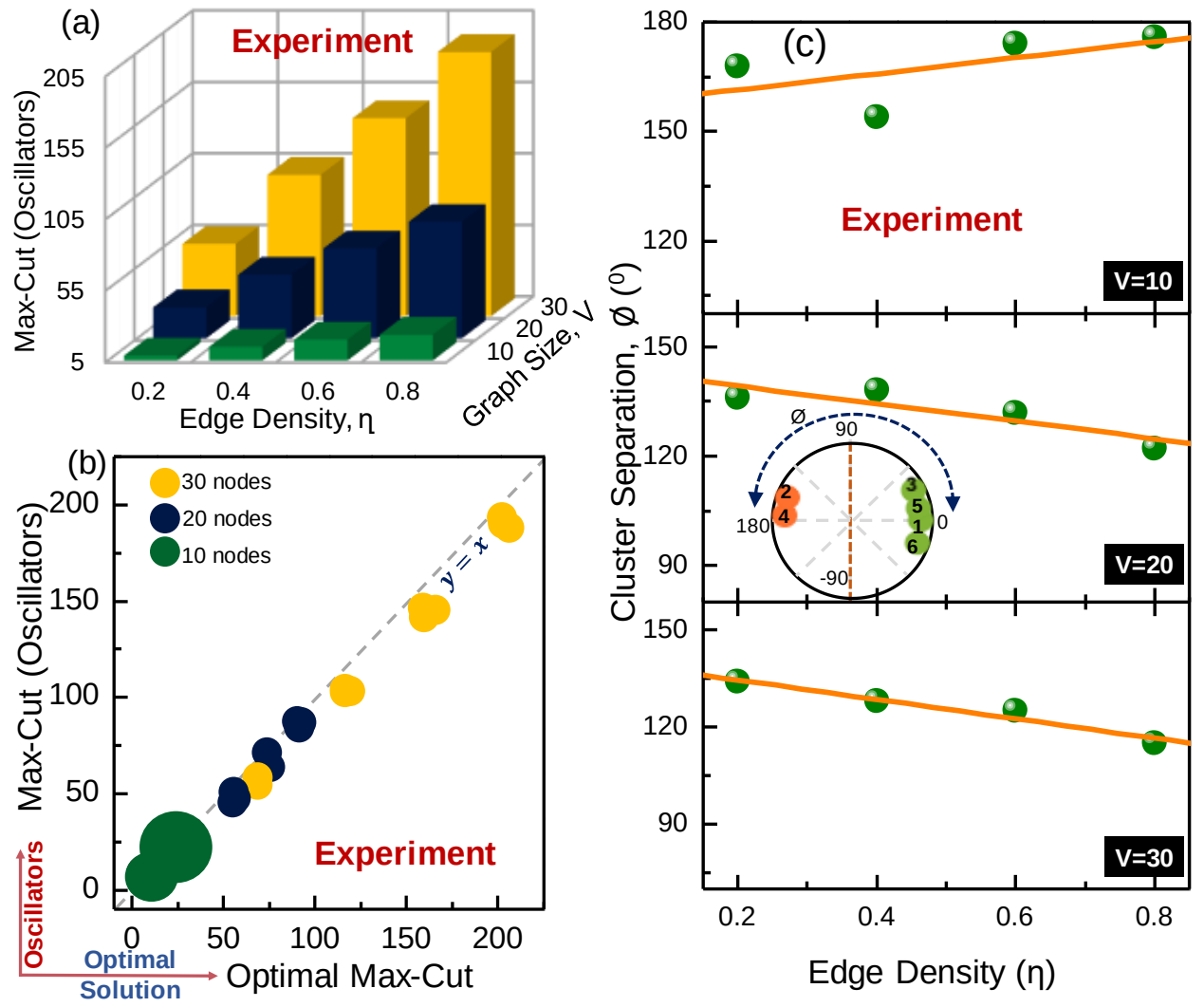

Fig. 2. (a) Bar plot showing the measured Max-Cut solutions for 36 randomly generated graph instances as a function of their size and edge density; (b) Bubble plot comparing (best case) Max-Cut solution obtained from the oscillators with the optimal Max-Cut of the graph; (c) Variation of cluster separation (i.e. angular separation between the two oscillator phases) with graph size and edge density.

We test our hardware on randomly generated graph instances with V=10, 20, 30 nodes, and having edge density, η= 0.2, 0.4, 0.6, 0.8 (η is the ratio of the number of edges in the graph to the number of edges in an all-to-all connected graph of same size); three graphs are tested for each combination of V and η (Fig. 2(a)) with each graph being measured 10 separate times. While the best solution has been considered in Fig. 2, the distribution of solutions over the 10 runs is shown in the supplement Fig. 1. It is evident that larger and denser graphs have larger Max-Cuts, and consequently, are more challenging to solve [1]. Fig. 2(b) shows a bubble plot comparing the value of the measured Max-Cut (best case) using the oscillators with optimal solution obtained using the BiqMac solver developed by Rendl *et al.* [1], [43] ; comparison of the mean value of the Max-Cut solution computed by the oscillators is shown in the supplement Fig. 2. It can be observed that the solution to most of the analyzed graphs lies near- or on the identity line (y=x) although larger graphs tend to show higher deviations from the optimal solution. As measured, the oscillator solution is within 99% of the optimal solution in 12 of the 36 graphs.

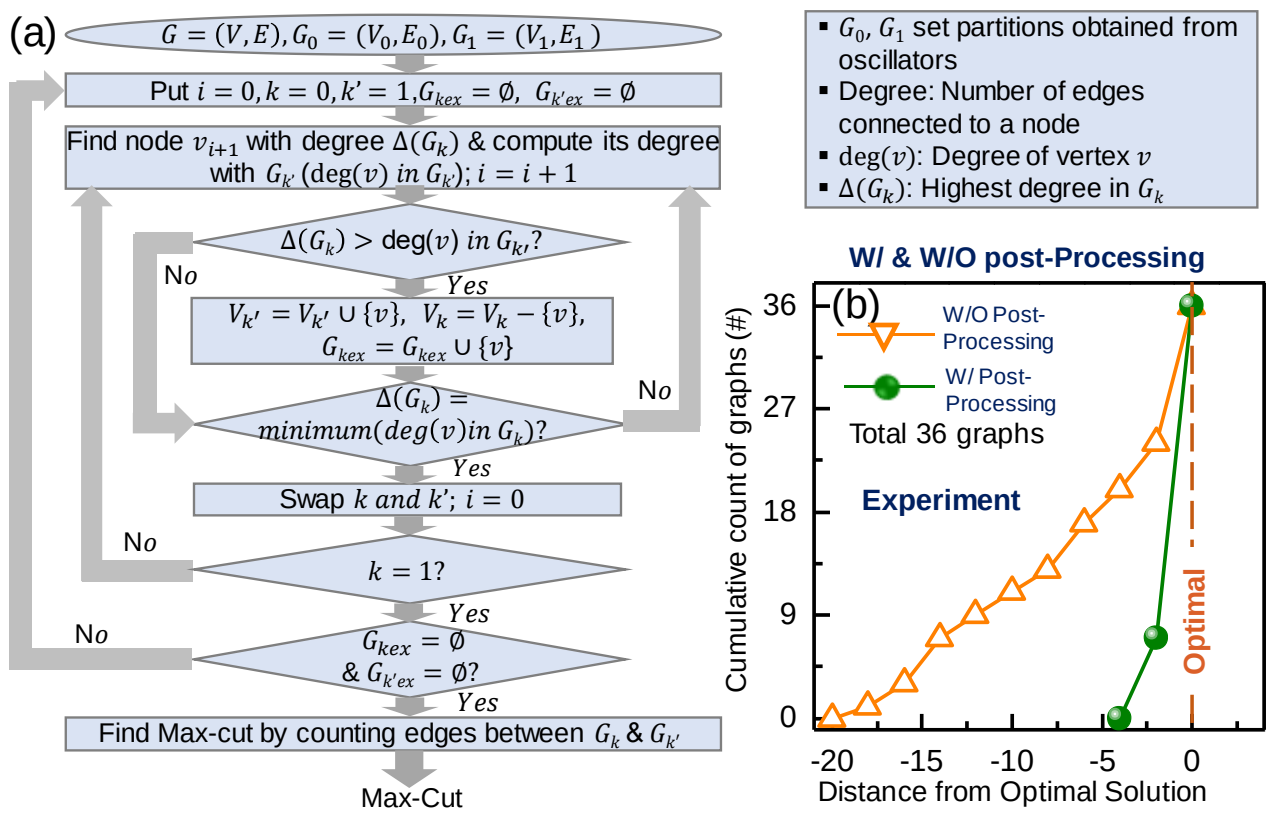

Fig. 3. (a) Flow chart for the local search-based post-processing scheme to improve the Max-Cut solution generated by the oscillators. (b) Deviation of the oscillator generated Max-Cut solution (w/ and w/o post-processing) from the optimal solution.

We note that the above experiments were performed without using a specialized annealing schedule, and the accuracy can be improved in the future by incorporating an annealing scheme [38]. Further, as described in the following section, the solution can be dramatically improved by using a simple polynomial-time local search scheme such that the solution is within the 99% of the optimal Max-Cut in 28 graphs, and equals the optimal Max-Cut in 26 of the 36 measured graphs. The larger value of the Max-Cut in the larger and denser graphs makes them challenging to solve. This property also manifests in the oscillator dynamics such as the cluster separation (defined as the difference between the mean phases of each cluster) shown in Fig. 2(c). Larger and denser graphs show reduced phase separation (i.e. more deviation from the ideal 180° phase difference) in comparison to smaller and sparser graphs implying that the system finds it increasingly challenging to attain the global energy minima corresponding to the optimal Max-Cut solution. Additionally, the effect of $V_{DD}$ variation and temperature are shown in the supplement Fig. 3.

## B. *Improving Measured Max-Cut Solution*

To improve the Max-Cut solution obtained from the oscillators, we explore a simple polynomial-time ($O(n^2)$) scheme based on local search as shown in the flow-chart in Fig. 3(a). Using the Max-Cut solution computed from the oscillators, the scheme proceeds by moving nodes between the sets if and only if the move increases the value of the Max-Cut. This process is repeated until no more nodes can be found that can increase the value of the cut. The cumulative graph count distribution as a function of the distance from optimal solution (i.e. difference between the Max-Cut solution obtained using the coupled oscillators and optimal Max-Cut) before and after post-processing shows the corresponding improvement in the solution for the experimentally measured graphs. The hardware-software approach produces the optimal Max-Cut in 26 (~72%) of the 36 graphs. Moreover, the oscillators are also effective in solving challenging dense graphs where they produce optimal solutions in 7 out of the 9 measured graphs with edge density η=0.8.

## C. *Comparison with other approaches*

Figure 4 shows a table comparing this work with other alternate approaches being explored to solve such computationally hard problems. While CMOS-based classical implementations do not reduce the fundamental complexity of the problem (as expected with quantum mechanical systems [41]), they offer a room temperature solution that can still provide a significant speed in comparison to digital computers owing to the inherent parallelism of the approach. Of these, coupled electronic oscillators provide a potentially promising low-power, integrated and compact solution. While larger coupled oscillator systems have been recently demonstrated [40], this work differentiates itself by exploring the computational dynamics of the oscillators in non-planar graphs with a wide range of edge densities, enabled through the implementation of an all-to-all reconfigurable capacitive coupling scheme.

| | **Hamerly *et al.* [41]** | **Takata *et al.* [44]** | **Takemoto *et al.* [20]** | **Su *et al.* [22]** | **Yamamoto *et al.* [21]** | **Chou *et al.* [38]** | **Dutta *et al.* [39]** | **Ahmed *et al.* [40]** | **This Work** |
|---|---|---|---|---|---|---|---|---|---|
| **Architecture** | Qubit (D-Wave) | Optical Parametric Oscillators | SRAM | Register | SRAM / Register | Coupled Oscillators (LC Oscillators) | Coupled Oscillators ($VO_2$ Oscillators) | Coupled Oscillators (Ring Oscillator) | Coupled Oscillators (Schmitt Trigger Oscillator) |
| **Coupling** | Nearest Neighbor | All-to-all | Nearest Neighbor | Nearest Neighbor | All-to-All | Discrete | Discrete | Hexagonally Connected Planar | All-to-All |
| **Type of Graphs Solved** | Chimera Graph | Cubic and Ring Graph | King's Graph | King's Graph | Graph with Arbitrary Connectivity | Graph with Arbitrary Connectivity | Graph with Arbitrary Connectivity | Planar Graph | Graph with Arbitrary Connectivity |
| **Maximum Graph Size Solved** | 2048 | 16 | 60K (2x30K) | 480 | 512 | 4 | 4 | 560 | 30 |
| **Operating Temperature** | 0.015 K | 300K | 300K | 300K | 300K | 300K | 300K | 300K | 300K |
| **Random Number Generator** | No | No | Yes | Yes | Yes | No | No | No | No |
| **Peak Power** | 25KW (Cryogenic Cooling) | 1 W (16 Spins) | Not Reported | 0.75 µW/Spin | 649mW (512 Spins) | Not Reported | Not Reported | 23mW (560 Spins) | 1.76mW (30 Spins) |
| **Total Measured Solutions** | N/A | N/A | Not Specified | Not Specified | Not Specified | 2000 | 1 | 1000 | 36 |

Fig. 4. Comparison of the present work with other implementations of the Ising model.

Finally, we would also like to point out that while the coupled oscillator-based approach is promising, system scalability will be a critical factor in deciding its eventual success. Scaled systems will require the design and implementation of specialized annealing schemes / schedules which can prevent the system from getting trapped in local minima and producing sub-optimal solutions; the role of annealing will be particularly critical in larger systems that have an increasingly complex solution space, and will be investigated in the future. Further, scaled systems are also likely to require optimization of the coupling architecture including additional design considerations such as managing the delay between the coupling elements. For instance, since practical graphs are unlikely to be very dense [45] the coupling architecture in larger systems could be optimized to have densely connected oscillator clusters with relatively sparse connectivity amongst them to achieve an optimal trade-off between functionality and reconfigurability.

## IV. Conclusion

In summary, we have experimentally investigated the computational properties of coupled relaxation oscillators to solve the NP-Hard Max-Cut problem by developing a prototype integrated IC of 30 relaxation oscillators with reconfigurable all-to-all coupling. Using minimal post-processing, we show that the oscillator-based approach computes high-quality approximate Max-Cut solutions even in non-planar graphs.

## References

[1] F. Rendl, G. Rinaldi, and A. Wiegele, “Solving Max-Cut to optimality by intersecting semidefinite and polyhedral relaxations,” *Math. Program.*, vol. 121, no. 2, pp. 307–335, 2010.

[2] J. C. A. D’Auriac, M. Preissmann, and A. S. Leibniz-Imag, “Optimal cuts in graphs and statistical mechanics,” *Math. Comput. Model*, vol. 26, no. 8–10, pp. 1–11, Oct. 1997.

[3] A. K. Hartmann, and M. Weigt, Phase Transitions in Combinatorial Optimization Problems: Basics, Algorithms and Statistical Mechanics. 2006.

[4] S. N. Dorogovtsev, A. V. Goltsev, and J. F. F. Mendes, “Critical phenomena in complex networks,” *Rev. Mod. Phys.*, vol. 80, no. 4, pp. 1275–1335, Oct. 2008.

[5] F. Liers, T. Nieberg, and G. Pardella. (2011). “Via Minimization in VLSI Chip Design Application of a Planar Max-Cut Algorithm.” [Online]. Available: http://e-archive.informatik.uni-koeln.de/630/

[6] J. J. Hopfield, and D. W. Tank, "“Neural” computation of decisions in optimization problems," *Biological cybernetics*, vol. 52, no. 3, pp. 141-152, July 1985.

[7] P. W. Shor, “Polynomial-time algorithms for prime factorization and discrete logarithms on a quantum computer,” *SIAM review*, vol. 41, no. 2, pp. 303-332, 1999.

[8] K. Roy, M. Sharad, D. Fan, and K. Yogendra, "Beyond charge-based computation: Boolean and non-Boolean computing with spin torque devices," in *International Symposium on Low Power Electronics and Design (ISLPED)*, pp. 139-142, Sept. 2013.

[9] A. Lucas, "Ising formulations of many NP problems," *Frontiers in Physics*, vol. 2, p. 5, Feb. 2014.

[10] C. Pan, and A. Naeemi, "Non-Boolean computing benchmarking for beyond-CMOS devices based on cellular neural network," *IEEE Journal on Exploratory Solid-State Computational Devices and Circuits*, vol. 2, pp. 36-43, Nov. 2016.

[11] F. L. Traversa, and M. Di Ventra, "Universal memcomputing machines," *IEEE transactions on neural networks and learning systems*, vol. 26, no. 11, pp. 2702-2715, Nov. 2015.

[12] S. Utsunomiya, K. Takata, and Y. Yamamoto, "Mapping of Ising models onto injection-locked laser systems," *Optics express*, vol. 19, no. 19, pp.18091-18108, 2011.

[13] M. W. Johnson *et al.*, “Quantum annealing with manufactured spins,” *Nature*, vol. 473, no. 7346, pp. 194–198, May 2011.

[14] A. D. King and C. C. McGeoch. (2014). “Algorithm engineering for a quantum annealing platform.” [Online]. Available: http://arxiv.org/abs/1410.2628

[15] V. N. Smelyanskiy *et al.* (2012). “A Near-Term Quantum Computing Approach for Hard Computational Problems in Space Exploration.” [Online]. Available: http://arxiv.org/abs/1204.2821

[16] Z. Wang, A. Marandi, K. Wen, R. L. Byer, and Y. Yamamoto, "Coherent Ising machine based on degenerate optical parametric oscillators," *Physical Review A*, vol. 88, no. 6, p. 063853, Dec. 2013.

[17] P. L. McMahon, A. Marandi, Y. Haribara, R. Hamerly, C. Langrock, S. Tamate, T. Inagaki, H. Takesue, S. Utsunomiya, K. Aihara, and R. L. Byer, "A fully programmable 100-spin coherent Ising machine with all-to-all connections," *Science*, vol. 354, no. 6312, pp.614-617, Nov. 2016.

[18] Y. Haribara, S. Utsunomiya, and Y. Yamamoto, “A coherent ising machine for MAX-CUT problems: Performance evaluation against semidefinite programming and simulated annealing,” *Lect. Notes Phys.*, vol. 911, pp. 251–262, 2016.

[19] M. Yamaoka, C. Yoshimura, M. Hayashi, T. Okuyama, H. Aoki, and H. Mizuno, “A 20k-spin Ising chip to solve combinatorial optimization problems with CMOS annealing,” *IEEE J. Solid-State Circuits*, vol. 51, no. 1, pp. 303–309, Dec. 2016.

[20] T. Takemoto, M. Hayashi, C. Yoshimura, and M. Yamaoka, "2.6 A 2× 30k-Spin Multichip Scalable Annealing Processor Based on a Processing-In-Memory Approach for Solving Large-Scale Combinatorial Optimization Problems," in *2019 IEEE International Solid-State Circuits Conference-(ISSCC)*, pp. 52-54, IEEE, 2019.

[21] K. Yamamoto *et al.*, "7.3 STATICA: A 512-Spin 0.25 M-Weight Full-Digital Annealing Processor with a Near-Memory All-Spin-Updates-at-Once Architecture for Combinatorial Optimization with Complete Spin-Spin Interactions," in *2020 IEEE International Solid-State Circuits Conference-(ISSCC)*, pp. 138-140, IEEE, 2020.

[22] Y. Su, H. Kim, and B. Kim, "31.2 CIM-Spin: A 0.5-to-1.2 V Scalable Annealing Processor Using Digital Compute-In-Memory Spin Operators and Register-Based Spins for

Combinatorial Optimization Problems," in *2020 IEEE International Solid-State Circuits Conference-(ISSCC)*, pp. 480-482, IEEE, 2020.

[23] G. Csaba, and W. Porod. "Coupled oscillators for computing: A review and perspective," *Applied Physics Reviews*, vol. 7, no. 1, p. 011302, Jan. 2020.

[24] D. E. Nikonov, P. Kurahashi, J. S. Ayers, H. J. Lee, Y. Fan, and I. A. Young. (2019). "A Coupled CMOS Oscillator Array for 8ns and 55pJ Inference in Convolutional Neural Networks." [online]. Available: https://arxiv.org/abs/1910.11803

[25] Z. Hull, D. Chiarulli, S. Levitan, G. Csaba, W. Porod, M. Pufall, W. Rippard *et al.*, "Computation with Coupled Oscillators in an Image Processing Pipeline," in *CNNA 2016; 15th International Workshop on Cellular Nanoscale Networks and their Applications*, Aug. 2016, pp. 1-2. VDE.

[26] A. Parihar, N. Shukla, M. Jerry, S. Datta, and A. Raychowdhury. "Vertex coloring of graphs via phase dynamics of coupled oscillatory networks," *Scientific reports*, vol. 7, no. 1, pp. 1-11, Apr. 2017.

[27] D. E. Nikonov, I. A. Young, and G. I. Bourianoff. (2014). "Convolutional networks for image processing by coupled oscillator arrays." [online]. Available https://arxiv.org/abs/1409.4469

[28] F. C. Hoppensteadt, and E. M. Izhikevich, "Synaptic organizations and dynamical properties of weakly connected neural oscillators," *Biological cybernetics*, vol. 75, no. 2, pp. 117-127, Aug. 1996.

[29] J. Torrejon, M. Riou, F. A. Araujo, S. Tsunegi, G. Khalsa, D. Querlioz, P. Bortolotti *et al.*, "Neuromorphic computing with nanoscale spintronic oscillators," *Nature*, vol. 547, no. 7664, pp. 428-431, Jul. 2017.

[30] G. Csaba, A. Papp, W. Porod, and R. Yeniceri, "Non-boolean computing based on linear waves and oscillators," in *45th European Solid State Device Research Conference*, IEEE, Graz, Austria, Sep. 2015, pp. 101-104.

[31] D. E. Nikonov, G. Csaba, W. Porod, T. Shibata, D. Voils, D. Hammerstrom, I. A. Young, and G. I. Bourianoff, "Coupled-oscillator associative memory array operation for pattern recognition," *IEEE Journal on Exploratory Solid-State Computational Devices and Circuits*, vol. 1, pp. 85-93, Nov. 2015.

[32] S. P. Levitan, Y. Fang, J. A. Carpenter, C. N. Gnegy, S. N. Janosik, S. Awosika-Olumo, D. M. Chiarulli, G. Csaba, and W. Porod, "Associative processing with coupled oscillators," in *International Symposium on Low Power Electronics and Design (ISLPED)*, IEEE, Sep. 2013, pp. 235-235.

[33] G. Csaba, A. Raychowdhury, S. Datta, and W. Porod, "Computing with coupled oscillators: Theory, devices, and applications," in *2018 IEEE International Symposium on Circuits and Systems (ISCAS)*, May 2018, pp. 1-5.

[34] S. P. Levitan., Y. Fang, D. H. Dash, T. Shibata, D. E. Nikonov, and G. I. Bourianoff, "Non-Boolean associative architectures based on nano-oscillators," in *13th International Workshop on Cellular Nanoscale Networks and their Applications*, Aug. 2012, pp. 1-6.

[35] G. Csaba, and W. Porod, "Computational study of spin-torque oscillator interactions for non-Boolean computing applications," *IEEE Transactions on Magnetics*, vol. 49, no. 7, pp. 4447-4451, Jul. 2013.

[36] T. Wang and J. Roychowdhury. (2017). "Oscillator-based Ising Machine." [Online]. Available: http://arxiv.org/abs/1709.08102

[37] T. Wang, L. Wu, and J. Roychowdhury, "New computational results and hardware prototypes for oscillator-based Ising machines," in *Proceedings of the 56th Annual Design Automation Conference 2019*, June 2019, pp. 1-2.

[38] J. Chou, S. Bramhavar, S. Ghosh, and W. Herzog, "Analog Coupled Oscillator Based Weighted Ising Machine," *Sci. Rep.*, vol. 9, no. 1, pp. 1–10, Oct. 2019.

[39] S. Dutta, A. Khanna, J. Gomez, K. Ni, Z. Toroczkai, and S. Datta, "Experimental Demonstration of Phase Transition Nano-Oscillator Based Ising Machine," in *Tech. Dig. - Int. Electron Devices Meet. IEDM*, Dec. 2019, pp. 911–914.

[40] I. Ahmed, P. W. Chiu, and C. H. Kim, "A Probabilistic Self-annealing Compute Fabric based on 560 Hexagonally Coupled Ring Oscillators for Solving Combinatorial Optimization Problems," in *Symposia on VLSI Technology and Circuits*, 2020.

[41] R. Hamerly *et al.*, "Experimental investigation of performance differences between coherent Ising machines and a quantum annealer," *Sci. Adv.*, vol. 5, no. 5, pp. 1–11, May 2019.

[42] S. W. Shin, G. Smith, J. A. Smolin, and U. Vazirani. (2014). "How 'Quantum' is the D-Wave Machine?" [Online]. Available: http://arxiv.org/abs/1401.7087

[43] Biq Mac Solver - Binary quadratic and Max cut Solver, http://biqmac.uni-klu.ac.at/, Aug 1, 2020.

[44] K. Takata *et al.*, "A 16-bit coherent Ising machine for one-dimensional ring and cubic graph problems," *Scientific reports*, vol. 6, p. 34089, 2016.

[45] G. Melancon, "Just how dense are dense graphs in the real world? A methodological note," *Proceedings of the 2006 AVI workshop on BEyond time and errors: novel evaluation methods for information visualization*, pp. 1-7, 2006.

# Supplementary Material

# Experimental Demonstration of a Reconfigurable Coupled Oscillator Platform to Solve the Max-Cut Problem

Mohammad Khairul Bashar*, Antik Mallick*, Daniel S Truesdell, Benton H. Calhoun, Siddharth Joshi, and Nikhil Shukla

This supplement includes:

(1) Statistical distribution of the Max-Cut solution computed by the oscillators

(2) Comparison of the mean value of the Max-Cut solution (obtained from the oscillators) with the optimal Max-Cut value

(3) Effect of $V_{DD}$ and temperature on the computational properties of the coupled oscillators

**1. Statistical distribution of the Max-Cut solution computed by the oscillators**

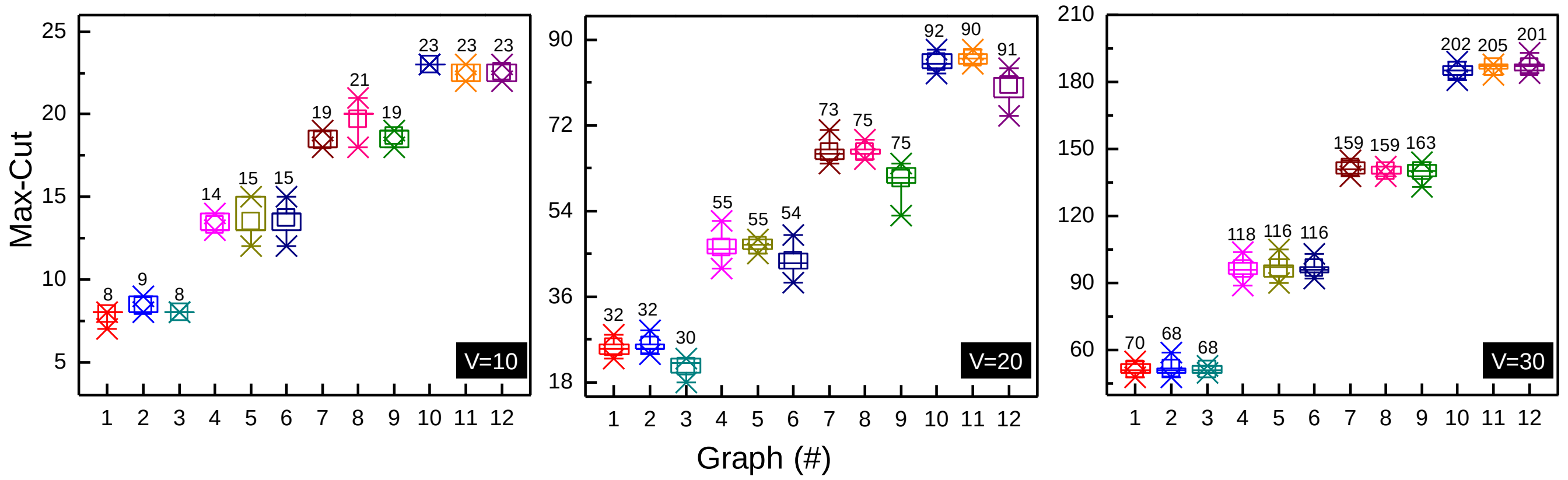

Fig. 1. Box plots showing the distribution of the Max-Cut computed by the oscillators for the 36 measured graphs. Each panel corresponds to graphs of different size (V). Three graphs are measured for each combination of V (=10, 20, 30) and η (=0.2, 0.4, 0.6, 0.8). The number besides the data point represents the optimal value of the solution.

Figure 1 shows box plots that detail the distribution of measured Max-Cut solution for the 36 measured graphs over 10 runs. Each panel consists of 12 graphs of a particular size (V=10,20, 30); three graphs are considered for each edge density value (η=0.2, 0.4, 0.6, 0.8).

**2. Comparison of Mean value of the Max-Cut solution with the optimal Max-Cut**

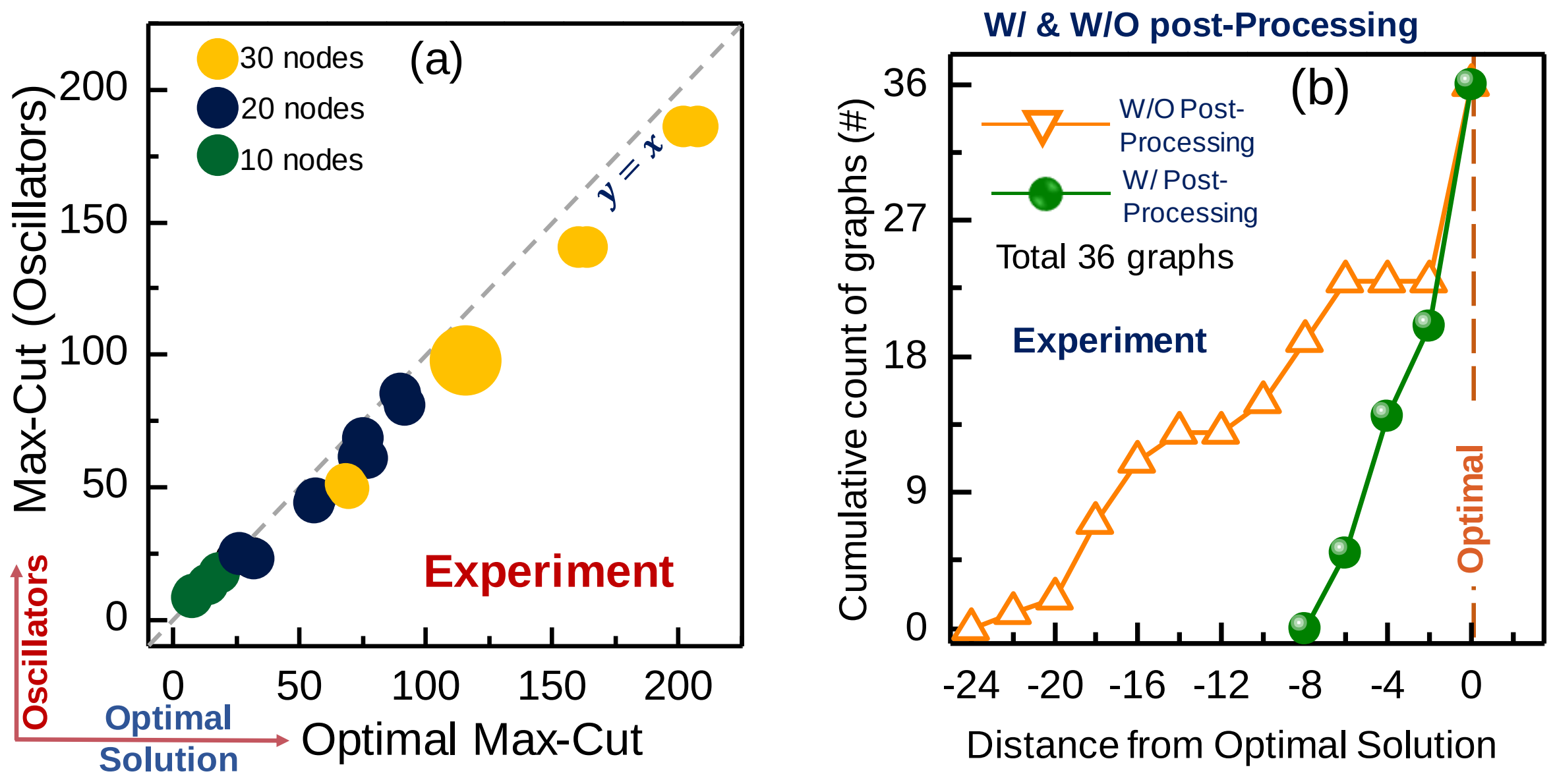

Fig. 2 (a) Bubble plot comparing mean value of (10 runs for each graph) Max-Cut solution obtained from the oscillators with the optimal Max-Cut. (b) Deviation of mean value of oscillator generated Max-Cut solution (w/ and w/o post-processing) from the optimal solution.

Fig. 2(a) shows the bubble plot comparing the mean value of the Max-Cut solution (averaged over 10 runs) with the optimal Max-Cut of the graph. Fig. 2(b) shows the cumulative graph count distribution as a function of distance from optimal Max-Cut before and after the post-processing. The mean value of the oscillator solution is within 95% of the optimal Max-Cut in 25 (70%) of the 36 measured graphs.

**3. Effect of $V_{DD}$ and temperature on the computational properties of the coupled oscillators**

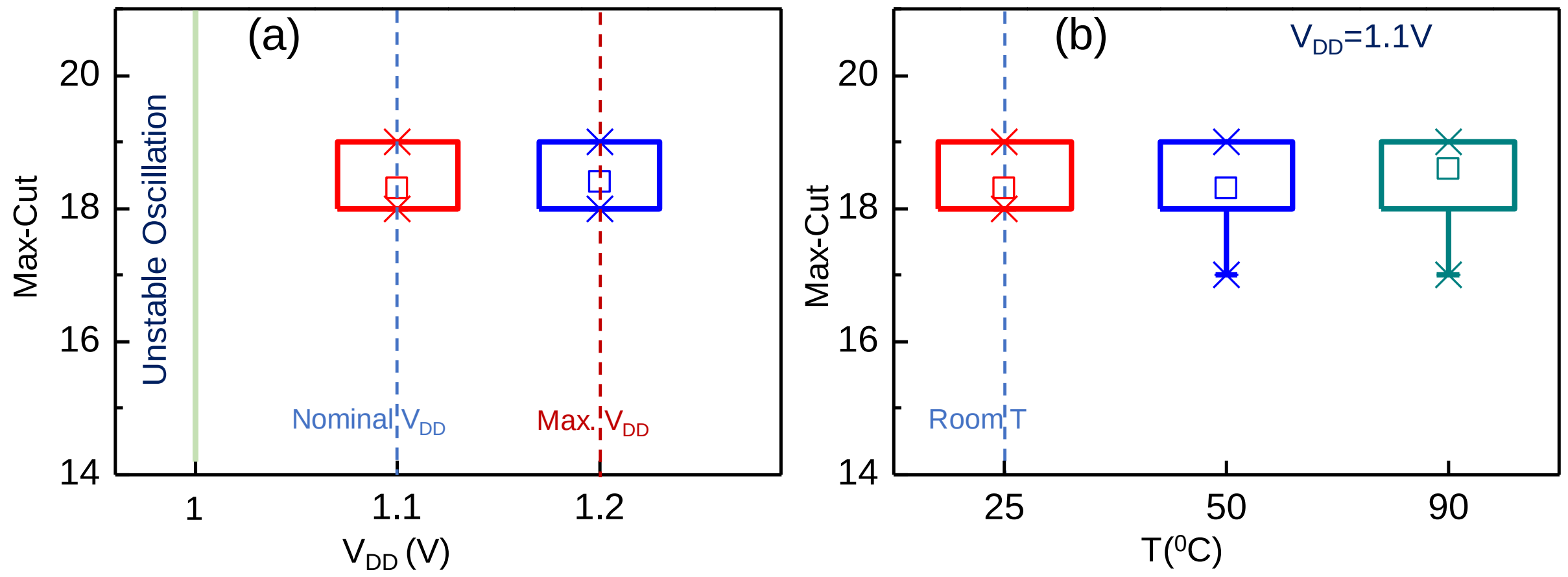

Fig. 3. Effect of (a) $V_{DD}$; (b) temperature on the computational properties of the coupled oscillators for a representative graph (V=10; η=0.6).

Fig. 3(a) shows the effect of $V_{DD}$ on the resulting Max-Cut solution for a representative graph (V=10; η=0.6). The nominal $V_{DD}$ used in this work is $V_{DD}$=1.1V. While the oscillators exhibit the phase bipartition at $V_{DD}$=1.2V, we observe that the oscillations become unstable below 1V. We believe that this is due to the change in operating (quiescent) point of the oscillators at lower bias. Fig. 3(b) shows the effect of temperature for the same graph.